# Dense electron-hole plasma in silicon light emitting diodes

**P D Altukhov and E G Kuzminov**

Ioffe Institute, Politekhnicheskaya Street 26, 194021 St. Petersburg, Russia

E-mail: pavel.altukhov@gmail.com

**Abstract.** Efficient electroluminescence of silicon light emitting p-n diodes with different sizes and shapes is investigated at room temperature. High quantum efficiency of the diodes, a long linear dependence of the electroluminescence intensity on the diode current and a low energy shift of the emission line in electroluminescence spectra with increasing diode current are explained by the self-compression of injected electron-hole plasma into dense electron-hole plasma drops. Experiments on space scanning of the electroluminescence intensity of the diodes support this conclusion. The plasma self-compression is explained by existence of an attraction in electron-hole plasma, compensating the plasma pressure. A decrease of the semiconductor energy gap due to a local lattice overheating, produced by the plasma, and the exchange-correlation interaction could contribute to this attraction. The self-focusing of the injection current can accompany the plasma self-compression.

## 1. Introduction

Physics and possible applications of efficient silicon light emitting diodes attract attention of a large number of researchers [1-14]. The maximum value of the quantum efficiency of silicon light emitting p-n diodes at room temperature is about 0.01 [7], and the quantum efficiency of band to band recombination radiation of ultrapure silicon can be about 0.1 [8]. At room temperature and low impurity concentration, the quantum efficiency of silicon light emitting diodes is $\eta = An/(w+An+Cn^2)$, where $A \approx 3\times10^{-15}$ cm$^3$s$^{-1}$ [15, 16] is the coefficient of radiative recombination, $w \approx 5\times10^4$ s$^{-1}$ is the probability of recombination via deep impurities, and $C \approx 2.3\times10^{-31}$ cm$^6$s$^{-1}$ is the coefficient of Auger recombination. The observed value $\eta \sim 0.01$ defines the density of injected electrons and holes in the region $n \sim (10^{17} \div 10^{18})$ cm$^{-3}$. This implies, that the injected electrons and holes are concentrated in small and dense plasma drops [1-3] with sizes of the drops much lower than sizes of the diodes. The condensation of injected carriers into dense electron-hole plasma drops is explained by the self-compression of injected carriers [1-3]. The plasma self-compression is caused by existence of an attraction in electron-hole plasma, compensating the plasma pressure [2]. A decrease of the semiconductor energy gap due to a local lattice overheating, produced by the plasma, gives an essential contribution to this attraction [2, 3]. Calculations of the exchange-correlation potential in electron-hole plasma in silicon at finite temperatures show, that the exchange-correlation interaction at the plasma density higher than $10^{17}$ cm$^{-3}$ is not negligible and also could give a contribution to reduction of semiconductor energy gap in silicon even at room temperature at low occupation numbers of electrons and holes [17, 18]. A reduction of the semiconductor energy gap, caused by the exchange-correlation interaction, was observed in highly doped silicon at the impurity concentration higher than $10^{17}$ cm$^{-3}$ at room temperature [19, 20].

We showed, that a negative heat capacity of electron-hole plasma at the temperatures $T \geq$ 300 K, concentration of the input diode power inside the plasma and weak diffusion of phonons at room temperatures could represent the main physical reasons and conditions of the plasma condensation [2]. Under these conditions, generation of phonons by the plasma results in a local overheating of the lattice and a reduction of the semiconductor energy gap inside the plasma. In such a way the self-organized potential well, attracting injected electrons and holes, is created [2]. At the negative heat capacity of the plasma, the plasma self-compression is accompanied by a decrease of plasma energy, observed as a low energy shift of the high energy edge of a recombination radiation line in electroluminescence spectra with increasing diode current [2, 3].

Here, we investigate efficient electroluminescence of silicon light emitting p-n diodes with different sizes and shapes and a surface distribution of the electroluminescence intensity of the diodes at room temperature. Two contributions to an attraction potential in electron-hole plasma are discussed: a reduction of the semiconductor energy gap due to a local lattice overheating, produced by the plasma, and the exchange-correlation potential. Presented results support our conclusion about the plasma self-compression in the silicon p-n diodes.

## 2. Results and discussion

Silicon p-n light emitting diodes were fabricated on n-type silicon substrates with a phosphorus concentration of about $10^{14}$ cm$^{-3}$. The p-n junction was formed by a surface layer, highly doped by boron under thermal diffusion, with a boron concentration of about $5\times10^{19}$ cm$^{-3}$. The diodes with circular and rectangular shapes and sizes equal to (1÷10) mm were used in the experiments. The thickness of the silicon layer, highly doped by boron, is about 1 micron. The Al strip metal layer with the thickness of the layer about 1 micron and the layer width about 0.3 mm was made on the emitting surface of the diodes. A surface layer, highly doped by phosphorus under thermal diffusion, with a phosphorus concentration of about $5\times10^{19}$ cm$^{-3}$ serves as a metal contact on the back side of the diodes. A Ni cover metal layer with a thickness of the layer of about 1 micron was made on the back side of the diodes. The thickness of the silicon substrates is 0.35 mm. The diode voltage of the diodes $V_g$ is applied between the top Al strip and the bottom Ni layer.

Under surface scanning experiments, the image of the emitting surface of the diode was focused by lenses on an input slit of a monochromator with the width and height of the slit equal to 0.1 mm. The recombination radiation was detected by a cooled photomultiplier using a photon counting system at a photon energy equal to 1.14 eV. The surface scanning of the electroluminescence intensity was produced by a step movement of the image along the slit with the space step equal to 0.1 mm along the x and y direction.

The emitting surface of the diode with rectangular shape is shown in figure 1.

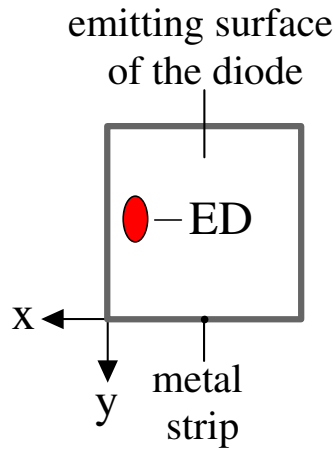

**Figure 1.** Emitting dots ED (one or two) of the silicon light emitting diode No. 3. The diode area is $4 \times 4$ mm$^2$.

The voltage-current dependences and the intensity-current dependences of silicon light-emitting diodes are shown in figure 2, and the electroluminescence spectra of these diodes are shown in figure 3. Presented spectra are corrected for the spectral response of the detected system. The spectral distribution of the observed TO emission line, corresponding to the band to band transitions, is correctly described by the formula $I = I_o \times exp(-E/kT') \times (E/kT')^2$, where $E = (h\nu - h\nu_o)$, $h\nu_o$ defines the photon energy quantum, corresponding to the low energy edge of the line, $T' = \xi T_e$ is the spectral temperature [2], $T_e$ is the electronic temperature of the electron-hole plasma. A difference between $T'$ and $T_e$ is explained by absorption of light in silicon plates [2]. In our p-n diodes this difference for the center of the emitting dots seems to be small, because recombination radiation was collected from the front surface of the diodes, and the light path in this case is small.

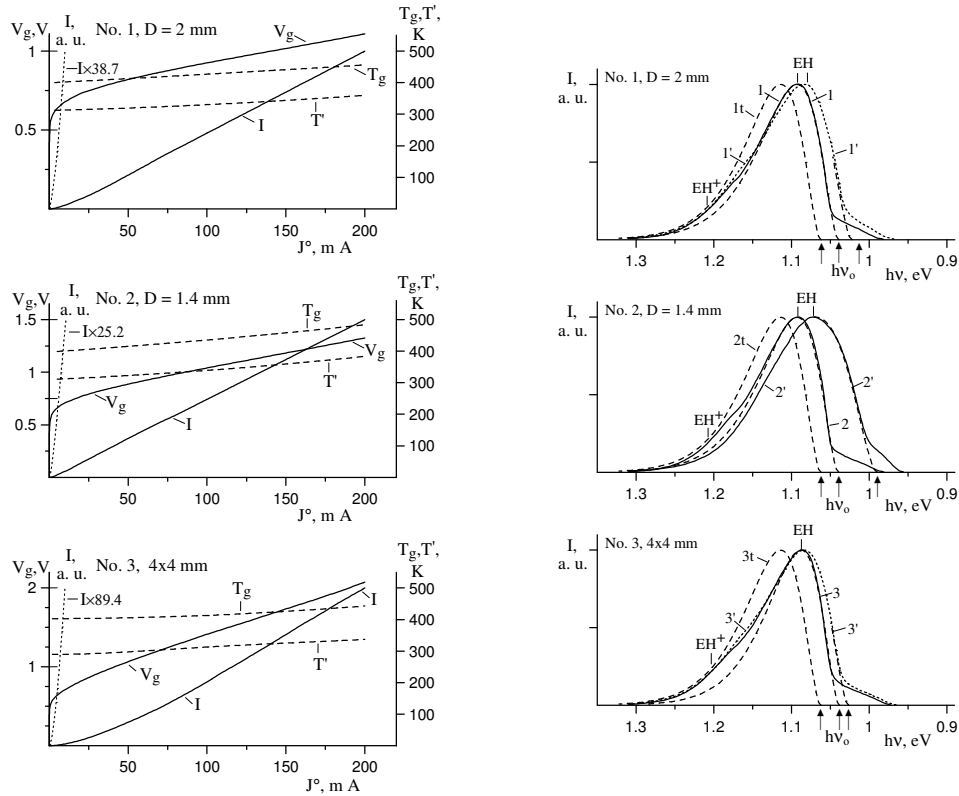

**Figure 2** (on the left side). Dependence of the diode voltage $V_g$, the electroluminescence intensity $I$, the lattice temperature $T_g$ inside the plasma [2], and the spectral temperature $T'$ on the diode current $J°$ for the circular silicon light emitting diodes No. 1, 2 with the diameter $D = 2$ mm and $D = 1.4$ mm and the rectangular diode No. 3 with the sizes $4 \times 4$ mm.

**Figure 3** (on the right side). Electroluminescence spectra of the silicon light emitting diodes (1, 1′, 2, 2′, 3, 3′) for the center of the emitting dots. The dashed lines 1t, 2t, 3t represent theory at the diode temperature $T = 300$ K. Other dashed lines represent theory for the corresponding electroluminescence spectra. Symbol EH indicates the TO-line, corresponding to emission of the TO-phonon under electron-hole recombination [2]. Symbol EH$^+$ indicates the TO-line, corresponding to absorption of the TO-phonon under electron-hole recombination [2]. The diode current is $J° = (2 \div 20)$ mA for the spectra 1, 2; $J° = 20$ mA for the spectrum 3; $J° = 200$ mA for the spectra 1′, 3′; $J° = 300$ mA for the spectrum 2′. The arrows $h\nu_o$ indicate the spectral position of the low energy edge of the TO-lines [2, 3].

The successful theoretical description of the line shape indicates absence of recombination radiation of excitons at the low energy edge of the line. It is explained by screening of excitons in dense electron-hole plasma at $n \geq 10^{17}$ cm$^{-3}$ [21]. Excitons could give a significant contribution to the recombination radiation line in the form of a pronounced low energy feature [16] only at low densities of injected carriers, defined as $n < 10^{16}$ cm$^{-3}$. Screening of excitons results in a decrease of the coefficient of radiative recombination due to a decrease of the contribution of excitons in electroluminescence spectra, and is also accompanied by a decrease of the electron-hole correlation enhancement factor and by a corresponding decrease of the coefficient of radiative recombination of free electrons and holes [22]. The spectral position of the low energy edge of the emission lines $h\nu_o = E_g - h\Omega_{TO}$ (figure 3) gives a value of the semiconductor energy gap $E_g$ inside the plasma and a value of the lattice temperature inside the plasma $T_g$ [2]. Here $h\Omega_{TO} = 58$ meV is the energy of the TO-phonon. The lattice temperature $T_g$, determined in such a way, should be close to the electronic temperature $T_e$ in our diodes, if the exchange-correlation interaction is weak. In this case $T_g$ can be used as the plasma temperature.

The quantum efficiency of the diodes is $\eta \approx (0.3 \div 0.5) \times 10^{-2}$. The same value of the quantum efficiency we measured previously for silicon p-n diode [3] and for tunneling silicon metal-oxide-semiconductor diodes [2]. We attribute the observed light emission to small dense electron-hole plasma drops [2], and results of a surface scanning of the electroluminescence intensity (figures 1 and 4) give us an opportunity for a conclusion about the self-compression of injected electron-hole plasma in the silicon p-n diodes.

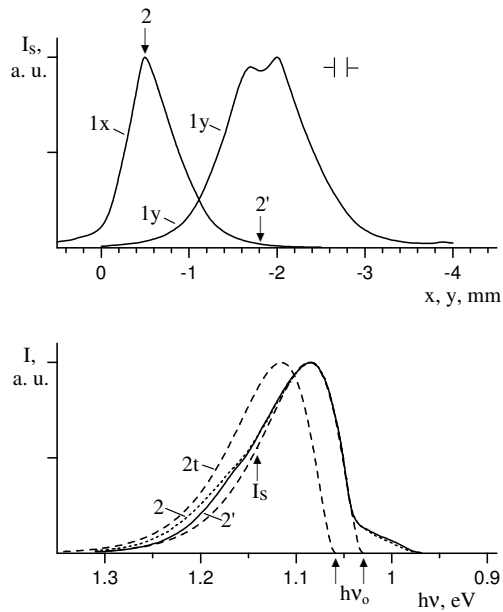

**Figure 4.** Surface distribution of the electroluminescence intensity $I_S$ (1x, 1y) at $h\nu = 1.14$ eV for the emitting dots ED and electroluminescence spectra (2, 2′) of the silicon p-n diode No. 3. x is the distance from the left edge of the diode in the x direction; y is the distance from the lower edge of the diode in the y direction (Figure 1). The diode current is $J^o = 0.2$ A, the diode temperature is $T = 330$ K, the temperature of the electron-hole plasma is $T_g = 430$ K. 1x: along the x direction at y = - 2 mm. 1y: along the y direction at x = - 0.5 mm. 2: at the maximum of the emitting dot at y = - 2 mm and x = - 0.5 mm. 2′: at y = - 2 mm, x = - 1.8 mm. 2t: theory for electron-hole gas at $T = 330$ K. The arrow $I_S$ at the spectrum 2′ indicates the photon energy for the space scanning experiments. The arrows $h\nu_o$ indicate the spectral position of the low energy edge of the TO emission lines [2, 3]. The arrows 2 and 2′ indicate the space position for the spectra 2 and 2′ respectively.

The strongly shifted to the low energy side of the spectrum emission lines of the diodes in comparison with the calculated emission line of rare electron-hole gas (figure 3, spectra 1t, 2t and 3t) represent evidence of an essential semiconductor energy gap reduction inside the plasma drops. The initial decrease of the semiconductor energy gap, observed at low diode currents, is equal to 23 meV. With increasing diode current the high energy edge of the emission line shifts to the low energy side of the spectrum as well as the low energy edge (figure 3). This demonstrates a decrease of the energy of the plasma drops with increasing diode current, proves the existence of the attraction potential, compensating the plasma pressure, and supports our conclusion about the plasma self-compression. Two contributions to the attraction potential can be essential: the semiconductor energy gap reduction, resulting from a local overheating, produced by the plasma [2], and the semiconductor energy gap reduction due to the exchange-correlation interaction [17, 18]. We can not divide these contributions, but it is supposed, that the first contribution dominates at high diode currents, producing the strong low energy shift of the emission line (figure 3). At low diode currents these contributions can be comparable. The role of the exchange-correlation interaction needs an additional investigation. If we neglected the exchange-correlation interaction, the temperature of the plasma drops $T_g$ could be calculated correctly by using of values of $h\nu_o$ [2]. The value of $T_g$, obtained under this assumption, is shown in figure 2.

We have found, that only one or two plasma drops can exist in our diodes (figures 1, 4). The diode current can be represented as $J^o = I/\eta = eAn^2V_d/\eta$, where $V_d$ is the volume of the injected plasma and $I$ is the radiative recombination current [2]. In accordance with this formula the long linear intensity-current dependences (figure 2) show, that the size of the plasma drops increases with increasing diode current, but the plasma density is constant, keeping a constant value of the quantum efficiency of the diodes. The non-linear part of the intensity-current dependences, observed at low diode currents, is narrower for diodes with lower sizes (figure 3). We assume, that the plasma density can be changed at the low diode currents in the region of this non-linear part.

The long linear intensity-current dependence of the plasma drops at room temperature reminds the linear dependence of recombination radiation of the electron-hole drops on the excitation intensity at low temperatures [23]. These electron-hole drops represent the degenerate electron-hole liquid. The degenerate electron-hole liquid in silicon can not be observed at room temperature, because the critical temperature of the electron-hole liquid in silicon is equal to 28 K [23], and the nature of the plasma drops at room temperature is different from the nature of the electron-hole liquid. The exchange-correlation attraction in the electron-hole liquid is sufficient to compensate the pressure, resulting from the kinetic energy. The exchange-correlation attraction can give a contribution to the reduction of the semiconductor energy gap in dense electron-hole plasma at room temperature [17, 18], but it is not sufficient itself to compensate the plasma pressure. A reduction of the semiconductor energy due to a local lattice overheating, produced by the plasma, gives an essential contribution to the attraction potential, necessary for compensation the plasma pressure, at the temperature higher than 250 K in the region of the negative heat capacity of the plasma [2]. The action of the resulting attraction in dense electron-hole plasma at room temperature is similar to the action of gravitation under formation of star matter. A self-heating of the electron-hole liquid, resulting from electron-hole recombination and producing instability of electron-hole drops and biexcitons, is possible at low temperatures in AgBr [24, 25]. But the observed bistability of luminescence [24, 25] does not indicate the local lattice temperature in the region of the negative heat capacity presumably due to high heat conductivity of the semiconductor at low temperatures.

The image of the emitting surface of the rectangular diode is represented by the brightly emitting dot, located at the center of an edge of the diode, and the weakly emitting area outside the dot (figures 1 and 4). This location of emitting dot is usual for most of the investigated diodes. But for some diodes one or two brightly emitting dot are located at corners of the diodes, and switching of the maximum intensity between diode corners can be observed [3]. The surface distribution of the electroluminescence intensity of the emitting dot is represented in figure 4 by the profile 1x for x direction and the profile 1y for y direction. The size of the emitting dot in the x direction is 0.6 mm. The profile 1y reveals a doublet structure, and we attribute the emitting dot to emission of two plasma drops. The surface profile width of the dot emission is defined by light propagation along silicon plates, depends on quality of silicon surface and can be higher than the plasma drop size. Multiple reflection and absorption of light in silicon plates influence the profile width as well as light emission from the diode influences the light path along the diode surface. Light propagation from the emitting dot along the silicon plate results in absorption of light in the high energy part of the spectrum and a corresponding narrowing of the emission line under measurements of the spectra at large distances from the emitting drop [3]. Light absorption is weak for light at the center of the emitting dot, resulting in a weak distortion of the emission spectra (spectra in figures 3, and spectrum 2 in figure 4). The observed difference between spectrum 2 and spectrum 2′ in the high energy part of the spectra in figure 4 is explained by a large light path for spectrum 2′. We estimate the maximum size of emitting plasma drops at high diode currents as 0.3 mm and the plasma density $n \geq 10^{17}$ cm$^{-3}$ in the plasma drops.

The observed shift of the emission line to the low energy side of the spectrum and the corresponding reduction of the semiconductor energy gap is not sufficient to explain all features of the plasma self-compression, especially, a significant difference between the carrier densities in the drops and outside the drops and a corresponding very weak emission intensity at a large distance from the plasma drops (figure 4). We assume, that the self-focusing of the injection current in the region of the plasma drops is an additional physical process, necessary for condensation of injected carriers into plasma drops. This self-focusing of the injection current can accompany the plasma self-compression presumably due to reduction of the semiconductor energy gap at the p-n junction in the region of plasma drops, resulting in a local decrease of the chemical potentials of electrons $\mu_e$ and holes $\mu_h$ at the n-type side of the p-n junction and a corresponding local enhancement of the injection current [3]. Sites of concentration of the injection current can be defined by the potential inhomogeneities. The potential inhomogeneities originate first of all from a finite resistance of the surface gate, formed by the highly doped silicon layer. The injection potential is equal to $(E_g - \mu_e - \mu_h)$. The injection potential should be maximum at diode corners, decreasing at the center of a diode edge. Nevertheless, the center of a diode edge is found to be a preferable site of the plasma drops, and nucleation phenomena for plasma drops have to be investigated to explain this result.

## 3. Conclusion

Our experimental results show, that dense electron-hole plasma represents the ground state of the system of injected carriers in silicon light emitting diodes at room temperature. An attraction in this system, compensating the electron-hole pressure, results in the self-compression of injected carriers into dense plasma drops. The plasma density and sizes of the plasma drops are estimated. The self-focusing of the injection current in the region of the plasma drops explains a very low density of injected carriers outside the plasma drops. It has been shown recently, that the temperature of injected carriers in a transistor is higher than the transistor temperature [26]. This supports our conclusion about an essential role of the local lattice overheating in the formation of dense plasma drops.